\documentclass[a4paper,11pt]{article}

\usepackage{tikz}
\usetikzlibrary{backgrounds}
\usepackage{caption}

\usepackage{indentfirst}

\usepackage{amsmath,amssymb}    %%Ò»Ð©ÊýÑ§·ûºÅ
\usepackage[T1]{fontenc} % if needed
\usepackage[textsize=footnotesize,textwidth=2.5cm]{todonotes}

\usepackage{comment}
\usepackage{mathrsfs}
\usepackage{color}
\usepackage{graphicx}
\usepackage{subfigure}
\usepackage{cite}               %%ÊÇÒýÓÃžñÊœ±äÆ¯ÁÁ
\usepackage{hyperref}           %%³¬ÁŽœÓ±ØÐë

\numberwithin{equation}{section}   %%¹«Êœ°ŽœÚ±àºÅ

\def \be {\begin{equation}}
\def \ee {\end{equation}}
\def \ba {\begin{array}}
\def \ea {\end{array}}
\def \bea{\begin{eqnarray}}
\def \eea{\end{eqnarray}}

\def \a {\alpha}

\def \D {\Delta}

\def \th {\theta}

\def \p {\partial}

\def \nn {\nonumber}

\def \hs {\hspace}

\begin{document}
\title{Supersymmetric Warped Conformal Field Theory}
\author{
Bin Chen$^{1,2,3}$,
Peng-Xiang Hao$^1$
and Yan-jun Liu$^1$
}
\date{}

\maketitle

\begin{center}
{\it
$^{1}$Department of Physics and State Key Laboratory of Nuclear Physics and Technology,\\Peking University, 5 Yiheyuan Rd, Beijing 100871, P.~R.~China\\
\vspace{2mm}
$^{2}$Collaborative Innovation Center of Quantum Matter, 5 Yiheyuan Rd, Beijing 100871, P.~R.~China\\
$^{3}$Center for High Energy Physics, Peking University, 5 Yiheyuan Rd, Beijing 100871, P.~R.~China}
\vspace{10mm}
\end{center}

\begin{abstract}

In this work, we study the supersymmetric warped conformal field theory in two dimensions. We show that the Hofman-Strominger theorem on symmetry enhancement could be generalized to the supersymmetric case. More precisely, we find that within a chiral superspace $(x^+,\th)$, a two-dimensional field theory with two translational invariance and a chiral scaling symmetry can have enhanced local symmetry, under the assumption that the dilation spectrum is discrete and non-negative. Similar to the pure bosonic case, there are two kinds of minimal models, one being $N=(1,0)$ supersymmetric conformal field theories, while the other being $N=1$ supersymmetric warped conformal field theories (SWCFT). We study  the properties of SWCFT, including the representations of the algebra, the space of states and the correlation functions of the superprimaries. 
%Symmetry plays an essential role in Modern Physics. Symmetry enhancement is particularly important and interseting in this community. D. M. Hofman and A. Strominger pointed out that a two dimensional field theory with two translational invariance and a chiral scaling symmetry, which indicate a $SL(2; R) \times U(1)$ global symmetry, can have enhanced local symmetry algebra\cite{Hofman:2011zj}. There are two minimal options for this algebra. One is CFT$_{2}$, and the other is Warped Conformal Field Theory(WCFT). In this paper, we generalize their results to a $N=2$ supersymmetric one, which also contains two minimal theories. One is the $N=(1,0)$ SCFT$_{2}$, the other is the $N=2$ supersymmetric Warped Conformal Field Theory(SWCFT). We studied the representation of $N=2$ SWCFT, including mode expansion of the algebra, space of the states and the transformations of the superfields. Then we calculated the two-point and three-point correlation functions of the $N=2$ SWCFT. Due to the infinite symmetries, the form of the correlation functions can be totally fixed.
\end{abstract}

\baselineskip 18pt
\thispagestyle{empty}

\newpage
\tableofcontents
%\newpage

%\noindent {\large{\bf Acknowledgments}}

 \section{Introduction}

  Symmetry plays an essential role in quantum field theories. The theories with more symmetries could be better constrained such that their dynamics might be investigated even nonperturbatively. For example, the supersymmetric field theories  have better UV behaviors, and the conformal invariant theories are expected to be solvable in the framework of conformal bootstrap. 
   
   In two dimensions (2D), the global symmetries in a quantum field theory with scaling symmetry could be enhanced. As shown by J. Polchinski in late 1980s\cite{Polchinski:1987dy}, a 2D Poincar\'e invariant QFT with scale invariance could become  conformal invariant, provided that  the theory is unitary and  the dilation spectrum is discrete and non-negative. In 2011, D. Hofman and A. Strominger \cite{Hofman:2011zj} relaxed the requirement of Lorentz invariance and studied the enhanced symmetries of the theory with chiral scaling. They obtained two kinds of minimal theories, 
 one being the two-dimensional conformal field theory (CFT$_2$)\cite{Belavin:1984vu} and  the other  being the so-called the warped conformal field theory (WCFT$_2$)\cite{Detournay:2012pc}. In a warped CFT$_2$, the global symmetry group is  $SL(2,R)\times U(1)$, and it is enhanced to an infinite-dimensional  group generated by an Virasoro-Kac-Moody algebra. Very recently, the symmetry enhancement in 2D QFT was generalized to the cases with global translations and anisotropic scaling symmetries\cite{Chen:2019hbj}. 
 In such 2D Galilean field theories with anisotropic scaling, the enhanced local symmetries are generated by the infinite dimensional spin-$\ell$ Galilean algebra with possible central extensions, under the assumption that the dilation operator is diagonalizable and has a discrete and non-negative spectrum. 
 
 WCFT$_2$ has  rich structures similar to CFT$_2$. Though they are not Lorentzian invariant, WCFT$_2$ shares the modular covariance like CFT$_{2}$. For finite temperature WCFT$_2$ defined on a torus, the modular property can be used to evaluate the density of states at high temperature, which gives a Cardy-like formula for the thermal entropy of WCFT$_2$\cite{Detournay:2012pc}.   Due to the infinite symmetries, WCFT$_2$ is highly constrained. The form of the two- and three-point functions are determined by the global warped conformal symmetry while the four-point functions can be determined up to an arbitrary function of the cross ratio\cite{Song:2017czq}.
 Specific models of WCFT$_2$ include chiral Liouville gravity\cite{Compere:2013aya}, free Weyl fermion\cite{Hofman:2014loa,Castro:2015uaa}, free scalars \cite{Jensen:2017tnb},and also the Sachdev-Ye-Kitaev models with complex fermions\cite{Davison:2016ngz,Chaturvedi:2018uov}.
For the study on other aspects of WCFT$_2$, see 
\cite{Castro:2015csg,Song:2016gtd,Azeyanagi:2018har,Apolo:2018eky,Chaturvedi:2018uov,Apolo:2018oqv,Song:2019txa}. 

 %Just like their CFT cousins, data of WCFTs consists of the spectrum of operators and the three-point function coefficients. A systematic and non-perturbative method which is possible to solve the theory is the conformal bootstrap program\cite{Polyakov:1974gs,Rattazzi:2008pe,Poland:2010wg,ElShowk:2012ht,El-Showk:2014dwa,Simmons-Duffin:2016gjk}.The idea is to use unitarity, associativity of the operator algebra and crossing symmetry to constrain the possible sets of consistent conformal field theories (CFTs). Similar to the conformal bootstrap, The warped conformal bootstrap program has been initialized \cite{Song:2017czq}.

%Entanglement is the most fundamental phenomenon in quantum mechanics. The entanglement entropy, has a wide
%range of applications in quantum information theory, condensed matter physics, general relativity, and even
%in high energy theory in recent years. The entanglement entropy and its related topics such as $R\acute{e}nyi$ Entanglement, Holographic Entanglement Entropy of WCFTs have been studied in\cite{Castro:2015csg,Song:2016gtd,Chen:2019xpb,Gao:2019vcc}. Interestingly, the entanglement entropy of WCFTs reveals hints of nonlocality along the U(1) direction, which is also observed in their correlation functions\cite{Song:2017czq}. Moreover, the holographic complexity, which is a new tool  to gain insight in the role of entanglement in the emergence of spacetime geometry in quantum gravity, has been also studied\cite{Auzzi:2018pbc,Auzzi:2019fnp}.

On the other hand, WCFT$_2$ plays an important role in the study of holography beyond the usual AdS/CFT correspondence. In \cite{Compere:2013bya}, it has been shown that under the Comp\`ere-Song-Strominger (CSS) boundary conditions, the asymptotic symmetry group of the AdS$_3$ gravity is generated by an Virasoro-Kac-Moody algebra. This leads to the conjecture that under the CSS boundary conditions, the AdS$_3$ gravity could be dual to a holographic warped conformal field theory. This AdS$_3$/WCFT correspondence has been studied in \cite{Song:2016gtd,Apolo:2018eky,Castro:2017mfj,Wen:2018mev,Chen:2019xpb,Lin:2019dji,Gao:2019vcc}. Moreover WCFT$_2$ could also appears in the WAdS$_3$/WCFT$_2$ correspondence\cite{Compere:2009zj,Aggarwal:2019iay,Ghodrati:2019bzz,Detournay:2019xgl}, in which the bulk gravity is a three-dimensional topological massive gravity.

%From holography point view, the infinite-dimensional algebra and the lack of Lorentz symmetry make WCFTs  useful in the study of holography beyond AdS spacetimes. In 
%WCFT originated from the search for holographic dual to a class of geometries with $SL(2,R)\times U(1)$ isometry, including warped AdS3 (WAdS) \cite{Anninos:2008fx} and the near horizon of extremal Kerr( NHEK) \cite{Bardeen:1999px,Guica:2008mu}. By studying the consistent asymptotical boundary conditions and corresponding asymptotic symmetry group gives a large class of gravitational theories are generated by a Virasoro-Kac-Moody algebra, including AdS3/WCFT\cite{Compere:2013bya},  WAdS/CFT \cite{Anninos:2008fx} WAdS/WCFT \cite{Compere:2009zj}, Kerr/CFT \cite{Guica:2008mu}, and so on. All these dualities which beyond the standard AdS/CFT correspondence, enable us a better understanding the true nature of the holographic phenomenon and the real world.

In this paper, we would like to generalize the study on  WCFT$_2$ to the supersymmetric case. We first study the supersymmetries on the warped flat geometry \cite{Hofman:2014loa} in two dimensions, which is essentially equivalent to a Newton-Cartan geometry\cite{Andringa:2010it,Christensen:2013lma,Jensen:2014aia,Bergshoeff:2014uea,Hartong:2014oma,Hartong:2014pma} with an additional scaling structure. The supersymmetrization could be done by including  Grassmannian coordinates into the bosonic directions to make warped ``superspace". However, it turns out that the minimal supersymmetry could be realized in a chiral  $N=(1,0)$ superspace. We then study the enhanced local symmetries, following the approach developed in\cite{Polchinski:1987dy} and \cite{Hofman:2011zj}. Just as in bosonic case, we find  two classes of minimal enhanced algebra. One generates the local symmetries  of $N=(1,0)$ SCFT$_2$, while the other one generates the symmetries of the supersymmetric warped conformal field theory (SWCFT$_2$). Furthermore, we discuss the radial quantization and the state-operator correspondence in SWCFT$_2$, analogous to the usual WCFT$_{2}$ case. We study the correlation functions of superprimaries in SWCFT$_2$ as well.  We notice that the correlation functions share the similar structure as the ones in the holomorphic sector of $N=(1,0)$ SCFT$_2$, with additional modifications from  $U(1)$ symmetry.

The remaining parts are organized as follows. In Section 2, we discuss the supersymmetries on the warped geometry  and set our notations. In Section 3, we generalize the Hofman-Strominger theorem to the supersymmetric case and show that  the global symmetries are enhanced to the local ones. In Section 4, we consider the Hilbert space and the representation of the NS sector of the SWCFT$_2$. After establishing the state-operator correspondence, we  discuss the transformations of the super-primaries. Then we calculate the two-point functions and three-point functions of the superprimary operators in the NS sector of the SWCFT, and discuss the higher-point functions. We conclude and give some discussions in Section 5. In Appendix, we discuss the conserved currents in the superspace and show that we can consistently work in the $N=(1,0)$ superspace.

%. Assuming that the dilation spectrum is discrete and non-negative, the supersymmetric theories coupled to Newton-Cartan geometry with global translation and scaling symmetries have infinitely conserved charges. This means

\section{Supersymmetries on Warped Geometry }
%\quad In this work, we use a superspace with two Bosonic coordinates $x^{a}=(x^{0},x^{1})$ and a single Grassmann coordinate $\theta$. We will consider local, unitary, translationally-invariant quantum field theories in this $2D$ superspace with a global scale invariance and a superymmetry. Here, the superymmetry is generated by a real supercharge. The transformation of coordinates under these symmetries are\footnote{here $a$, $b$ are Bosonic constants and $\epsilon$ is a Grassmann constant.}
Let us start from a two-dimensional unitary local field theory with translational invariance and a chiral scaling symmetry. The transformation of coordinates under these symmetries are
\begin{equation}\label{sym1}
x^{a}\to x^{a}+\delta^{a},\qquad  x^{a}\to {\lambda^{a}}_{b}x^{b},
\end{equation}
where ${\lambda^{a}}_{b}$ is a scaling matrix:
\begin{equation}
{\lambda^{a}}_{b}=\begin{pmatrix} \lambda & 0 \\ 0 & 1 \end{pmatrix}
\end{equation}
As shown in \cite{Hofman:2011zj},  the theory would have enhanced local symmetries. 
There are two kinds of minimal theories. One kind is the two-dimensional conformal field theory (CFT$_{2}$), while the other kind is the two-dimensional warped conformal field theory (WCFT$_2$). %We now want to add supersymmetries and see what the story will be. We have to describe fermionic representations in both minimal theories. Because of CFT$_{2}$ will be enhanced to SCFT$_{2}$, which is well known in the literature, let us mainly consider the WCFT case. 
For WCFT$_2$, in addition to the symmetries (\ref{sym1}), there is   a generalized boost symmetry 
\begin{equation}
x^{a}\to {\Lambda^{a}}_{b}x^{b}
\end{equation}
where ${\Lambda^{a}}_{b}$ is the boost matrix
\begin{equation}
{\lambda^{a}}_{b}=\begin{pmatrix} 1 & 0 \\ v & 1 \end{pmatrix}.
\end{equation}

The WCFT$_2$ can be defined consistently in a warped geometry, which is a variant of the Newton-Cartan geometry with an additional scaling structure\cite{Detournay:2012pc}. In the warped geometry, there are one vector and one one-form 
\begin{equation}
q^{a}=\begin{pmatrix} 0  \\ 1 \end{pmatrix},\qquad q_{a}=(1,0),
\end{equation}
which are invariant under the boost. And 
there is an antisymmetric tensor $h_{ab}$ which is also invariant under the boost, 
\begin{equation}
h_{ab}=\begin{pmatrix} 0 & 1 \\ -1 & 0 \end{pmatrix},\qquad h^{ab}=-h_{ab=}\begin{pmatrix} 0 & -1 \\ 1 & 0 \end{pmatrix}. 
\end{equation}
$h_{ab}$ can be used to lower  the indices, but one should keep in mind that $h_{ab}$ is not the metric of the warped geometry.

In the warped geometry, one may define the fermionic representations. The first step is to consider the gamma matrix algebra. The gamma matrix algebra is given  by the warped Clifford algebra
\begin{equation}
\big\{\Gamma^{a},\Gamma^{b} \big\}=2q^{a}q^{b},
\end{equation}
where the gamma matrices are:
\begin{equation}
\Gamma^{0}=\begin{pmatrix} 0 & 0 \\ 1 & 0 \end{pmatrix},\qquad \Gamma^{1}=\begin{pmatrix} 1 & 0 \\ 0 & -1 \end{pmatrix}.
\end{equation} 
 The lower-index gamma matrices are defined  by
\begin{equation}
\Gamma_{a}=h_{ab}\Gamma^{b}.
\end{equation}
This definition proves quite useful as it allows us to define the boost generator as
\begin{equation}
\overline{B}=\frac{1}{8}h_{ab}[\Gamma^{a},\Gamma^{b}].
\end{equation}
One can check that it acts on the gamma matrices as they are in a vector representation
\begin{equation}
[\overline{B},\Gamma^{c}]=q_{a}\Gamma^{a}q^{c},\quad \text{or}\quad[\overline{B},\Gamma_{c}]=-q^{a}\Gamma_{a}q_{c}.
\end{equation}
The operators generating the translations  will be donated by $H_{a}=(H_{0},H_{1})=(H,\overline{P})$, and they are of course in a vector representation of the boost generator
\begin{equation}
[\overline{B},H^{c}]=q_{a}H^{a}q^{c},\quad \text{or}\quad[\overline{B},H_{c}]=-q^{a}H_{a}q_{c}.
\end{equation} 
The two-dimensional spinor space is spanned by $\Psi_{0}$, $\Psi_{1}$ as follows
\begin{equation}
\Psi_{A}=\begin{pmatrix} \Psi_{0} \\ \Psi_{1} \end{pmatrix} .
\end{equation}
From this it is easy to see that
\begin{equation}
\overline{B}\Psi_{0}=0,\qquad \overline{B}\Psi_{1}=\frac{1}{2}\Psi_{0}.
\end{equation}
The definition for the dual representation is
\bea
\overline{\Psi}^{A}=\epsilon^{AB}\Psi_{B}=(\Psi^{0},\Psi^{1})=(\Psi_{1},-\Psi_{0}).
\eea
where the $\epsilon^{AB}$ is given by
\begin{equation}
\epsilon^{AB}=-\epsilon_{AB}=\begin{pmatrix} 0 & 1 \\ -1 & 0 \end{pmatrix}
\end{equation}
One can easily show that the quantity $\overline{\Psi}\Psi$ is a scalar under the boost.

Now, let us introduce the supercharge operator $Q=(Q_{0},Q_{1})^{T}$. The commutators of supercharges are
\begin{equation}
i\big\{Q_{A},\overline{Q}^{B}\big\}=2{(\Gamma^{a}H_{a})_{A}}^{B}.
\end{equation}
They can be written in terms of the component operators
\bea
i\big\{Q_{1},Q_{1}\big\}=2H,\hs{3ex}
i\big\{Q_{0},Q_{1}\big\}=2\overline{P},\hs{3ex}
i\big\{Q_{0},Q_{0}\big\}=0.
\eea
From these commutators, one can easily find that $H$ and $Q_{1}$ are superpartners  under the action of $Q_{1}$, so are  $\overline{P}$ and $Q_{0}$.

For simplicity, we will donate $Q_{1}$ by $Q_{+}$ and $Q_{0}$ by $Q_{-}$ in the following discussion. Moreover, we will donate $x^{a}=(x^{0},x^{1})$ by $x^{a}=(x^{+},x^{-})$, and the Grassmannian coordinates by
\begin{equation}
\theta^{A}=(\theta^{+},\theta^{-}) .
\end{equation}

%Then we can see for a tensor $T_{\mu}$, its components in light coordinates is:
%\begin{equation}
%T_{+}=\frac{1}{2}(-T_{0}+T_{1}),\qquad  T_{-}=\frac{1}{2}(T_{0}+T_{1}),
%\end{equation}
%$\gamma $ matrices are :
%\begin{equation}
%\gamma^{\beta}=(i\sigma^{2},\sigma^{1}),
%\end{equation}
%where $\sigma^{1}$ and $\sigma^{2}$ are two of the three Pauli matrices and satisfy $\gamma_{\alpha}\gamma_{\beta}+\gamma_{\beta}\gamma_{\alpha}=\eta_{\alpha\beta}$.
A general superfield is  defined on the superspace, and can be expanded as a power series in $\theta^{+}$ and $\theta^{-}$
\begin{equation}
\Phi(\mathbf{z})\equiv \Phi(x^{+},x^{-},\theta^{+},\theta^{-})=A_{1}(x^{+},x^{-})+\theta^{+}A_{2}(x^{+},x^{-})+\theta^{-}A_{3}(x^{+},x^{-})+\theta^{+}\theta^{-}A_{4}(x^{+},x^{-}).
\end{equation}
The transformation of any field $\Phi$ under the generator $G$ is given by
\begin{equation}
\delta_{\epsilon}\Phi=i[\epsilon G,\Phi].
\end{equation}

In the appendix, we discuss the conserved charges of the theory in the superspace and their corresponding supercurrents. We find that there exists a minimal superspace  in which the right-moving supersymmetry can be turned off consistently. As we show in the next section, even only with the supersymmetry in the left-moving sector, the right-moving global symmetry gets enhanced and supersymmetrized as well. 

\section{Enhanced Symmetries}

In this section, we study the enhanced symmetry of two dimensional quantum field theory, whose global symmetry is generated by the left-moving  translation $H$, the dilation $D$, the right-moving translation $\overline{P}$ and the supersymmetries $Q_{+}$. We will work in the chiral superspace $(x^+, x^-, \th^+)$. For simplicity, we denote $\th^+=\th$. Now that for the global charges their related supercurrents   depend only on one Grassmannian  coordinate $\theta$, we will discuss the  enhanced symmetry in this $N=1$ chiral superspace. As in \cite{Hofman:2011zj}, we  assume that the eigenvalue spectrum  of $D$ is discrete and non-negative and there exists a complete basis of $N=1$ local superfields $\Phi_{i}$. 

A general superfield can be expanded as  
\be
\Phi(x^{+},x^{-},\th)=\varphi (x^{+},x^{-})+\theta \psi (x^{+},x^{-}).
\ee
It satisfy:
\bea
i[H,\Phi_{i}]&=&\partial_{+}\Phi_{i},\qquad i[\overline{P},\Phi_{i}]=\partial_{-}\Phi_{i},\\
i[D,\Phi_{i}]&=&(x^{+}\partial_{+}+\frac{\theta}{2}\partial_{\theta}+\lambda_{i})\Phi_{i},\\
i[\epsilon Q_{+},\Phi_{i}]&=&\epsilon(\partial_{\theta}-\theta\partial_{+})\Phi_{i},\\
i[\epsilon Q_{-},\Phi_{i}]&=&\epsilon(-2\theta\partial_{-})\Phi_{i},
\eea
where $\lambda_{i}$ is the superweight of $\Phi_{i}$ and $\int_{C}d\Phi_{i}=0$ for any closed contour $C$. The translational plus the dilational invariance and the supersymmetry  restrict the form of the vacuum two-point functions  
\begin{equation}
\langle \Phi_{i}(x^{+},x^{-},\theta)\Phi_{j}(x^{+\prime },x^{-\prime },\theta^{\prime})\rangle=\frac{f_{ij}(x^{-}-x^{-\prime })}{(x^{+}-x^{+\prime }-\theta\theta^{\prime})^{\lambda_{i}+\lambda_{j}}},
\end{equation}
where $f_{ij}$ are some unknown functions.

\subsection{From left global symmetries to local symmetries}
%\subsubsection{Enhancement of Left Translation Symmetries}

The global charges $H$, $D$, $\overline{P}$ are associated to  the supercurrents $\mathbb{H}$, $\mathbb{D}$ and $\mathbb{\overline{P}}$, respectively. All of these supercurrents have  shift freedom \cite{Hofman:2011zj}, which can be used to ``gauge" the currents to satisfy the canonical commutation relations
\bea
i[H,\mathbb{H}_{\pm}]&=&\partial_{+}\mathbb{H}_{\pm}   ,\qquad    i[H,\mathbb{\overline{P}}_{\pm}]=\partial_{+}\mathbb{\overline{P}}_{\pm},\notag\\
i[H,\mathbb{D}_{\pm}]&=&\partial_{+}\mathbb{D}_{\pm}-\mathbb{H}_{\pm}.\\
i[\overline{P},\mathbb{H}_{\pm}]&=&\partial_{-}\mathbb{H}_{\pm} ,\qquad     i[\overline{P},\mathbb{\overline{P}}_{\pm}]=\partial_{-}\mathbb{\overline{P}}_{\pm},\notag\\
i[\overline{P},\mathbb{D}_{\pm}]&=&\partial_{-}\mathbb{D}_{\pm}.
\eea
This implies that $\mathbb{H}_{\pm}$, $\mathbb{\overline{P}}_{\pm}$ are local operators, but $\mathbb{D}_{\pm}$ must have explicit dependence on the $x^{+}$ coordinate. 
The weights of the global charges (\ref{diemnsion1})-(\ref{diemnsion2}) imply
\bea
i[D,\mathbb{H}_{+}]&=x^{+}\partial_{+}\mathbb{H}_{+}+\frac{\theta}{2}\partial_{\theta}\mathbb{H}_{+}+\frac{3}{2}\mathbb{H}_{+},\nn\\
i[D,\mathbb{H}_{-}]&=x^{+}\partial_{+}\mathbb{H}_{-}+\frac{\theta}{2}\partial_{\theta}\mathbb{H}_{-}+\frac{1}{2}\mathbb{H}_{-},\nn\\
i[D,\mathbb{\overline{P}}_{+}]&=x^{+}\partial_{+}\mathbb{\overline{P}}_{+}+\frac{\theta}{2}\partial_{\theta}\mathbb{\overline{P}}_{+}+\frac{1}{2}\mathbb{\overline{P}}_{+},\nn\\
i[D,\mathbb{\overline{P}}_{-}]&=x^{+}\partial_{+}\mathbb{\overline{P}}_{-}+\frac{\theta}{2}\partial_{\theta}\mathbb{\overline{P}}_{-}-\frac{1}{2}\mathbb{\overline{P}}_{-},\nn\\
i[D,\mathbb{D}_{+}]&=x^{+}\partial_{+}\mathbb{D}_{+}+\frac{\theta}{2}\partial_{\theta}\mathbb{D}_{+}+\frac{1}{2}\mathbb{D}_{+},\nn\\
i[D,\mathbb{D}_{-}]&=x^{+}\partial_{+}\mathbb{D}_{-}+\frac{\theta}{2}\partial_{\theta}\mathbb{D}_{-}-\frac{1}{2}\mathbb{D}_{-}.
\eea

The $\mathbb{D}_{\pm}$'s have explicit coordinate dependence. Let us  write the current in terms of local operators as in \cite{Hofman:2011zj}. Defining $\mathbb{S}_{\pm}$ by
\begin{equation}\label{A}
\mathbb{D}_{\pm}=x^{+}\mathbb{H}_{\pm}+\theta\mathbb{S}_{\pm}.
\end{equation}
One can easily find that
\begin{equation}
 i[H,\mathbb{S}_{\pm}]=\partial_{+}\mathbb{S}_{\pm},\qquad i[\overline{P},\mathbb{S}_{\pm}]=\partial_{-}\mathbb{S}_{\pm},
\end{equation}
and
\begin{equation}
 i[D,\mathbb{S}_{+}]=x^{+}\partial_{+}\mathbb{S}_{+}+\mathbb{S}_{+},\qquad i[D,\mathbb{S}_{-}]=x^{+}\partial_{+}\mathbb{S}_{-}.
\end{equation}
So we conclude $(\mathbb{S}_{+},\mathbb{S}_{-})$ are local operators of weight $(1,0)$.\\

The conservations of the dilation current and left-translation current yield
\bea
 \partial_{+}\mathbb{D}_{-}+\partial_{-}\mathbb{D}_{+}&=&x^{+}(\partial_{+}\mathbb{H}_{-}+\partial_{-}\mathbb{H}_{+})+\theta(\partial_{+}\mathbb{S}_{-}+\partial_{-}\mathbb{S}_{+})+\mathbb{H}_{-}\notag\\&=&0,
\eea
which leads to 
\bea
\mathbb{H}_{-}=-\theta(\partial_{-}\mathbb{S}_{+}+\partial_{+}\mathbb{S}_{-}).
\eea
Then we use the shift freedom in the currents to shift away $\mathbb{S}_{+}$
\begin{equation}
\mathbb{H}_{\pm}\to \mathbb{H}_{\pm}\pm \theta\partial_{\pm}\mathbb{S}_{+},\qquad \mathbb{D}_{\pm}\to \mathbb{D}_{\pm}\pm \theta\partial_{\pm}(x^{+}\mathbb{S}_{+}).
\end{equation}
One can check that the commutators and  the conservations of the currents  remain consistent. Now, the equation (\ref{A}) becomes
\begin{equation}
\mathbb{D}_{+}=x^{+}\mathbb{H}_{+},\qquad \mathbb{D}_{-}=x^{+}\mathbb{H}_{-}+\theta\mathbb{S}_{-},
\end{equation}
and
\begin{equation}
\mathbb{H}_{-}=-\theta\partial_{+}\mathbb{S}_{-}.
\end{equation}
Because of $\mathbb{S}_{-}$ is a local operator of weight zero, then from the general form of the two-point function, we have 
\begin{equation}
 \langle \mathbb{S}_{-}\mathbb{S}_{-}\rangle=f_{\mathbb{S}_{-}}(x^{-}),
\end{equation}
which implies
\begin{equation}\label{null}
\mathbb{H}_{-}=0.
\end{equation}
Assuming 
\bea
\mathbb{H}_{+}(x^{+},x^{-},\theta)=&h_{0}(x^{+},x^{-})+2\theta h_{1}(x^{+},x^{-}), 
\eea
the conservation law and the equation (\ref{null}) yield
\bea
\mathbb{H}_{+}(x^{+},x^{-},\theta)=&h_{0}(x^{+})+2\theta h_{1}(x^{+}).
\eea
This fact immediately leads to the existence of two sets of conserved charges. Defining
\begin{equation}
 T_{\xi}=-\frac{1}{2\pi}\int dx^{+}d\theta \xi(x^{+},\theta)\mathbb{H}_{+},
\end{equation}
where $\xi=\frac{1}{2}\alpha(x^{+})+\theta a(x^{+})$ with $\a$ and $a$ being the function of $x^+$,  we have
\bea
T_{0a}=-\frac{1}{2\pi}\int dx^{+} a(x^{+}) h_{0},\\
T_{1\alpha}=-\frac{1}{2\pi}\int dx^{+}\alpha(x^{+})h_{1}.
\eea
There is another set of conserved charges
\begin{equation}
 \overline{J}_{\chi}=\frac{1}{2\pi}\int dx^{-}\chi(x^{-})\mathbb{S}_{-},
\end{equation}
which may lead to other symmetries \cite{Hofman:2011zj,Hofman:2014loa}. 
As we are going to discuss the minimal algebra, we choose $\mathbb{S}_{-}=0$.
%In \cite{Hofman:2011zj} it was explained that the operator $s_{-}$ in their case could be responsible for yet another U(1) Kac-Moody symmetry. In \cite{Hofman:2014loa} was shown that for the minimal cases corresponding to CFTs and WCFTs this can't be the case. We may only discuss the minimal algebra in this work, so we choose $\mathbb{S}_{-}=0$.

%\begin{equation}
% \overline{B}:x^{-}\to x^{-}+vx^{+}
%\end{equation}
%This symmetry is generalized boost symmetry between $x^{+}$ and $x^{-}$.

The algebra spanned by the conserved bosonic charges $T_{1\alpha}$ has been done in  \cite{Hofman:2011zj}:
\begin{equation}\label{super1}
i[T_{1\alpha},T_{1\beta}]= T_{1(\alpha^{\prime}\beta-\alpha\beta^{\prime})}+\frac{c_{0}}{48\pi}\int dx^{+}(\alpha^{\prime\prime}\beta^{\prime}-\alpha^{\prime}\beta^{\prime\prime}),
\end{equation}
where the prime denotes the derivative with respect to $x^+$, $\alpha^{\prime}\equiv \partial_{+}\alpha$. This is the same as the algebra of the left-moving conformal generators on the Minkowski plane with the central charge $c_{0}$.

%\subsubsection{Enhancement of Supersymmetry $Q_{+}$}
Let us now work out the algebra spanned by adding the fermionic charges $T_{0a}$. The global charges are
\bea
H&=&-\frac{1}{4\pi}\int dx^{+}d\theta  1\cdot \mathbb{H}_{+}=-\frac{1}{2\pi}\int dx^{+}h_{1},\\
D&=&-\frac{1}{4\pi}\int dx^{+}d\theta\theta  x^{+}\cdot \mathbb{H}_{+}=-\frac{1}{2\pi}\int dx^{+}x^{+}h_{1},\\
Q_{+}&=&-\frac{1}{2\pi}\int dx^{+}d\theta\theta  \mathbb{H}_{+}=-\frac{1}{2\pi}\int dx^{+}h_{0}.
\eea
The actions of $H$ and $D$ on $\mathbb{H}_{+}$ imply
\bea
i[H,T_{0a}]&=&-T_{0a^{\prime}},\\
i[D,T_{0a}]&=&-T_{0(\frac{a}{2}-a^{\prime}x^{+})}.
\eea
This in turn implies that the action of $T_{0a}$ on $h_{1}$ is
\begin{equation}
 i[h_{1},T_{0a}]=-\frac{3}{2}a^{\prime}h_{0}-\frac{1}{2}a\partial_{+}h_{0}+\partial_{+}^{2}O_{a}.
\end{equation}
Furthermore we have
\begin{equation}
i [T_{1\alpha},T_{0a}]=T_{0(\frac{\alpha^{\prime }a}{2}-\alpha a^{\prime})}-\frac{1}{2\pi}\int dx^{+}\alpha \partial_{+}^{2}O_{a}.
\end{equation}
The scaling symmetry plus the locality imply that $O_a$ must be of the form $O_a=c_{1}a$ with $c_{1}$ being a local operator of weight $\frac{1}{2}$. But the Jacobi identity with the third operator $T_{1\beta}$ implies that $c_{1}=0$. So we arrive at
\begin{equation}\label{super2}
 i[T_{1\alpha},T_{0a}]=T_{0(\frac{\alpha^{\prime }a}{2}-\alpha a^{\prime})}.
\end{equation}

Next, the action of $Q_{+}=T_{01}$ on $h_{0}$ is $i\{T_{01},h_{0}\}=2h_{1}$. This implies $i\{T_{01},T_{0a}\}=2T_{1a}$ and hence
\begin{equation}
i \{T_{0a},h_{0}\}=2ah_{1}+\partial_{+}g_{a}, 
\end{equation}
where $g_{a}$ is to be determined. Integrating both sides with  $-\frac{1}{2\pi} dx^{+}b(x^{+})$ gives
\begin{equation}
i \{T_{0a},T_{0b}\}=2T_{1ab}-\frac{1}{2\pi}\int dx^{+}b\partial_{+}g_{a}.
\end{equation}
The scaling symmetry and the exchange symmetry under  $a\leftrightarrow b$ imply $g_{a}=c_{2}a^{\prime}$, where  $c_{2}$ is a constant number. The Jacobi identity with the third operator $T_{1\alpha}$ implies that $c_{2}=\frac{c_{0}}{3}$. Then
\begin{equation}\label{super3}
 i\{T_{0a},T_{0b}\}=2T_{1ab}+\frac{c_{0}}{6\pi}\int dx^{+}a^{\prime}b^{\prime}.
\end{equation}
We recognize the equations (\ref{super1}), (\ref{super2}), (\ref{super3}) as the superconformal algebra on the Minkowski plane with the central charge $c_{0}$.

\subsection{From right global symmetries to local symmetries}

In general, $\mathbb{\overline{P}}_{\pm}$ can be written in form of 
\bea
\mathbb{\overline{P}}_{+}=p_{0}(x^{+},x^{-})+2\theta p_{1}(x^{+},x^{-}),\\
\mathbb{\overline{P}}_{-}=p_{3}(x^{+},x^{-})+2\theta p_{2}(x^{+},x^{-}).
\eea
%As in the paper \cite{Hofman:2011zj}, $p_{1}=p_{1}(x^{+})$ and $p_{2}=p_{2}(x^{-})$. Supersymmetry requires $p_{0}=p_{0}(x^{+})$ and $p_{3}=0$, so $p_{-}$ is singlet under the supersymmetry.\\ 
The fact that $\mathbb{\overline{P}}_{-}$ is a local superfield of weight $-\frac{1}{2}$  implies that $p_{2}$ is a weight-zero local field. From the two-point function of $p_{2}$, we get $\partial_{+}p_{2}=0$. The current conservation then implies $\partial_{-}p_{1}=0$. It follows that
\begin{equation}
p_{1}=p_{1}(x^{+}),\qquad p_{2}=p_{2}(x^{-}).
\end{equation}
 The supersymmetry requires $p_{0}=p_{0}(x^{+})$ and $p_{3}=0$, hence $p_{-}$ is a singlet under the supersymmetry.
Now we have
\bea
\mathbb{\overline{P}}_{+}&=&p_{0}(x^{+})+2\theta p_{1}(x^{+}),\\
\mathbb{\overline{P}}_{-}&=&p_{2}(x^{-}).
\eea

In the case $\mathbb{\overline{P}}_{+}=0$, we have infinitely many charges given by
\begin{equation}
\overline{T}_{1\alpha}=\frac{1}{2\pi}\int dx^{-}\alpha(x^{-})p_{2}.
\end{equation}
The algebra spanned by $\overline{T}_{1\alpha}$ gives the right-moving Virasoro algebra on Minkowski plane \cite{Hofman:2011zj}. In this case, the enhanced local symmetry is generated by the left-moving super-Virasoro algebra and the right-moving Virasoro algebra. It gives the local symmetry of $N=(1,0)$ SCFT$_2$. 
%\subsubsection{Enhancement of Right Translation Symmetries}

In the case \ $\mathbb{\overline{P}}_{-}=0$, we have infinitely many left-moving charges
\begin{equation}
J_{\eta}=-\frac{1}{2\pi}\int dx^{+}d\theta\eta(x^{+},\theta)\mathbb{\overline{P}}_{+},
\end{equation}
where $\eta=\frac{1}{2}\eta(x^{+})+\theta c(x^{+})$. Then
\bea
 J_{0c}&=-\frac{1}{2\pi}\int dx^{+}d\theta\theta c \mathbb{\overline{P}}_{+}=-\int dx^{+}c p_{0},\\
 J_{1\eta}&=-\frac{1}{2\pi}\int dx^{+}d\theta\cdot \eta \mathbb{\overline{P}}_{+}=-\int dx^{+}\eta p_{1}.
\eea
The bosonic sector of the algebra are simply\cite{Hofman:2011zj}
\bea 
 i[J_{1\eta},J_{1\chi}]&=&-\frac{k}{8\pi}\int dx^{+}(\chi^{\prime}\eta-\chi\eta^{\prime}),\label{current1}\\
  i[T_{1\alpha},J_{1\eta}]&=&-J_{1\alpha\eta^{\prime}}.
\eea
The equation (\ref{current1}) is a $U(1)$ Kac-Moody current algebra and the constant $k$ parameterizes the central element.

To find the fermionic sector of the enhanced symmetry, we need to consider other commutators. Firstly we study the  commutator $[T_{0a},J_{1\eta}]$. Note that the action $Q_{+}=T_{01}$ on $\mathbb{\overline{P}}_{+}$ implies
\begin{equation}
i[T_{01},J_{1\eta}]=\frac{1}{2} J_{0\eta^{\prime}}.
\end{equation}
This in turn implies that the action of $J_{1\eta}$ on $h_{0}$ is
\begin{equation}\label{eq1}
[h_{0},J_{1\eta}]=\frac{1}{2}\eta^{\prime}p_{0}+\partial_{+} O_{0\eta}.
\end{equation}
The scaling symmetry plus the locality imply that $O_{0\eta}$ must be of the form $O_{0\eta}=c_{3}\eta$, where $c_{3}$ is a local operator of weight $\frac{1}{2}$. Consider the zero mode of $J_{1\eta}(\eta=1)\equiv J_{11}$, which act as $\partial_{-}$, we  have $i[J_{11},h_{0}]=0=\partial_{+}c_{3}$. This leads to the fact that $c_3$ must be independent of $x^+$. On the other hand, $c_{3}$ is an operator of weight $\frac{1}{2}$ under the chiral scaling,  we conclude that $c_{3}$ must be zero. Now integrating both sides of the equation (\ref{eq1}) with $-\frac{1}{2\pi} dx^{+}a$ gives
\begin{equation}
i[T_{0a},J_{1\eta}]=\frac{1}{2}J_{0(a\eta^{\prime})}.
\end{equation}

%\subsubsection{Enhancement of Supersymmetry $Q_{-}$}
Let us now work out the algebra spanned by $J_{0c}$. Due to the fact that the zero mode $J_{11}$ acts as $\partial_{-}$, we have $i[J_{11},p_{0}]=0$. This implies $i[J_{11},J_{0c}]=0$ and hence 
\begin{equation}
i[J_{0c},p_{1}]=\partial_{+}X_{c}.
\end{equation}
Again, the scaling symmetry plus the locality imply $X_{c}=0$, then
\begin{equation}\label{current2}
i[J_{0c},J_{1\eta}]=0.
\end{equation}

We also need the  commutator $[T_{1\alpha}, J_{0c}]$. The action of $H$ on $p_{0}$ implies $i[H,J_{01}]=0$, which in turn implies $i[h_{1},J_{01}]=\partial_{+}(p_{0}+Y)$ with $Y$ a local operator of weight $\frac{1}{2}$. Integrating both sides with  $-\frac{1}{2\pi}dx^{+}\alpha$ gives $i[T_{1\alpha}, J_{01}]=\frac{1}{2\pi}\int dx^{+}\alpha^{\prime}(p_{0}+Y)$. This gives
\begin{equation}
i[T_{1\alpha}, p_{0}]=-\alpha^{\prime}(p_{0}+Y)+\partial_{+}Z_{\alpha}.
\end{equation}
The scaling symmetry plus the locality implies $Z_{\alpha}$ must be of the form $Z_{\alpha}=c_{4}\alpha$ with $c_{4}$ being a local operator of weight $\frac{1}{2}$. The action of $D_1$ on $p_{0}$ implies that $Y=-\frac{p_{0}}{2}$ and $c_{4}=p_{0}$, hence
\begin{equation}
i[T_{1\alpha}, p_{0}]=\alpha p_{0}^{\prime}+\frac{\alpha^{\prime}p_{0}}{2}.
\end{equation}
Then we have
\begin{equation}
i[T_{1\alpha}, J_{0c}]=-J_{0(\alpha c^{\prime}+\frac{\alpha^{\prime}c}{2})}.
\end{equation}

Next we turn to the  anti-commutator $\{T_{0a}, J_{0c}\}$. The fact $i\{Q_{+},p_{0}\}=2p_{1}$ implies $i\{Q_{+},J_{0c}\}=2J_{1c}$. This in turn implies  
\begin{equation}
i\{h_{0}, J_{0c}\}=2cp_{1}+\partial_{+}O_{c}.
\end{equation}
The scaling symmetry plus the locality imply $O_{c}$ must be of form $O_{c}=c_{5}c$ with $c_{5}$ being a local operator of weight zero.  After integrating both sides with $-\frac{1}{2\pi} dx^{+}a(x^{+})$, we get
\begin{equation}
i\{T_{0a}, J_{0c}\}=2J_{1(ac)}-\frac{c_{5}}{2\pi}\int dx^{+}ac^{\prime}.
\end{equation}
The Jacobi identity with the third operator $T_{1\alpha}$ implies $c_{5}=0$. Finally we have
\begin{equation}
i\{T_{0a}, J_{0c}\}=2J_{1(ac)}.
\end{equation}

Finally we consider the  anti-commutator $\{J_{0c},J_{0d}\}$. The scaling symmetry implies that $\{J_{01},p_{0}\}=c_{6}$, where $c_{6}$ must be a weight-zero constant number. This implies $\{J_{01},J_{0c}\}=-c_{6}\int dx^{+}c$, which in turn gives
\begin{equation}
i\{p_{0},J_{0c}\}=c_{6}c+\partial_{+}W_{c}.
\end{equation}
Again, the scaling symmetry plus the locality imply that $W_{c}=0$. At last, we get
\begin{equation}\label{current3}
i\{J_{0c},J_{0d}\}=-c_{6}\int dx^{+}cd.
\end{equation}
The appropriate normalizations of $J_{0c}, J_{1\eta}$ can always help us to set $c_{6}=-\frac{k}{4\pi}$, then we recognize (\ref{current1}), (\ref{current2}), (\ref{current3}) as the $U(1)$ super-Kac-Moody algebra (SKMA) on the Minkowski plane with the central charge $k$. 

\subsection{Mode expansion}
 The supersymmetric Virasoro-Kac-Moody algebra consists of a 
super-Virasoro algebra
\bea
i[T_{1\alpha},T_{1\beta}]&= &T_{1(\alpha^{\prime}\beta-\alpha\beta^{\prime})}+\frac{c_{0}}{48\pi}\int dx^{+}(\alpha^{\prime\prime}\beta^{\prime}-\alpha^{\prime}\beta^{\prime\prime}),\\
i[T_{1\alpha},T_{0a}]&=&T_{0(\frac{\alpha^{\prime }a}{2}-\alpha a^{\prime})},\\
i\{T_{0a},T_{0b}\}&=&2T_{1ab}+\frac{c_{0}}{6\pi}\int dx^{+}a^{\prime}b^{\prime}, 
\eea
a super-Kac-Moody algebra
\bea
 i[J_{1\eta},J_{1\chi}]&=&-\frac{k}{8\pi}\int dx^{+}(\chi^{\prime}\eta-\chi\eta^{\prime}),\\
i[J_{0c},J_{1\eta}]&=&0,\\
i\{J_{0c},J_{0d}\}&=&\frac{k}{4\pi}\int dx^{+}cd, 
\eea
and the semidirect product of the super-Virasoro and super-Kac-Moody algebras
\bea
i[T_{1\alpha},J_{1\eta}]&=&-J_{1\alpha\eta^{\prime}},\\
i[T_{0a},J_{1\eta}]&=&\frac{1}{2} J_{0(a\eta^{\prime})},\\
i[T_{1\alpha}, J_{0c}]&=&-J_{0(\alpha c^{\prime}+\frac{\alpha^{\prime}c}{2})},\\
i\{T_{0a}, J_{0c}\}&=&2J_{1(ac)}.
\eea

Let us put the theory on a cylinder and find the mode expansion of the above algebra. The coordinate transformation is 
\begin{equation}
x^{+}=e^{i\phi}.
\end{equation}
Using the new coordinate $\phi$, we choose test functions $\alpha_{n}=(x^{+})^{n+1}=e^{i(n+1)\phi}$, $a_{r}=e^{i(r+\frac{1}{2})\phi}$, $\eta_{n}=e^{in\phi}$ and $c_{r}=e^{i(r-\frac{1}{2})\phi}$, where $n\in \mathbb{Z}$ and $r\in \mathbb{Z}+\frac{1}{2}$ for the Neveu-Schwarz(NS) sector or $r\in \mathbb{Z}$ for the Ramond(R) sector. Letting $L_{n}=iT_{1\alpha_{n}}$, $G_{r}=iT_{0a_{r}}$, $P_{n}=J_{1\eta_{n}}$ and $S_{r}=J_{0c_{r}}$, then the commutation relations in terms of the charges $\{L_{n}, P_{m},G_{r},S_{s}\}$ are as follows\footnote{We have set $c_{0}=c$ and $\delta_{n+m}=\delta_{n+m,0}$.}. The 
super-Virasoro algebra is generated by 
\be
[L_{m},L_{n}]=(m-n)L_{m+n}+\frac{c}{12}m(m^{2}-1)\delta_{m+n},\ee
\bea
[L_{m},G_{r}]&=&(\frac{m}{2}-r)G_{m+r},\\
\{G_{r},G_{s}\}&=&2L_{r+s}+\frac{c}{3}(r^{2}-\frac{1}{4})\delta_{r+s}.
\eea
The super-Kac-Moody algebra is generated by 
\be
 [P_{m},P_{n}]=\frac{k}{2}m\delta_{m+n},\hs{3ex}
[P_{m},S_{r}]=0,\hs{3ex}
\{S_{r},S_{s}\}=\frac{k}{2}\delta_{r+s}.
\ee
The semi-direct product part of the super-Virasoro and super-Kac-Moody algebras is generated by\footnote{we have substitute $-P$ for $P$.}
\be
[L_{m},P_{n}]=-nP_{m+n},\hs{3ex}
[G_{r},P_{m}]=-\frac{m}{2} S_{m+r},\ee
\be
[L_{m}, S_{r}]=-(\frac{m}{2}+r)S_{m+r},
\hs{3ex}
\{G_{r}, S_{s}\}=2P_{r+s}.
\ee

These algebras are the same as those appearing in the supersymmetric Wess-Zumino-Witten(SWZW) model \cite{McArthur:1989yd,DiVecchia:1984nyg,Fuchs:1986ew,Kiritsis:1986wx}. We would like to stress that the super-Kac-Moody algebra in the SWZW model is generated by internal symmetries, while here it is generated by the symmetric transformations in the superspace in SWCFT$_2$. The algebra of the SWZW model consists of two copies   (the left-moving and the right-moving) of  algebras, while there is only one copy (left-moving) of the algebra in  SWCFT$_2$.

Another remarkable point is on the supersymmetry in the right-moving sector. Our construction starts from the left-moving superspace, but the right-moving sector gets supersymmetrized as well. This could be understood from the diffeomorphism of WCFT$_2$. Recall that the diffeomorphism of WCFT$_2$ is generated by 
\be
x^+ \to f(x^+), \hs{3ex}x^-\to x^-, 
\ee
and 
\be
x^+\to x^+, \hs{3ex}x^-\to x^-+g(x^+)
\ee
The diffeomorphism could be generalize to chiral superspace such that the supersymmetry in the left-moving sector is transferred to the right-moving sector. Actually, the superconformal transformation in the left-moving sector  is\cite{Polchinski:1998rr}  
\bea
{x^{+}}^{\prime}&=&f(x^{+})+\theta F(x^{+}),\\
{\theta}^{\prime}&=&\phi(x^{+})+\theta p(x^{+}).
\eea
Here $f(x^+), p(x^+)$ are holomorphic functions and $F(x^+),\phi(x^+)$ are anticommuting holomorphic functions, satisfying the following relations
 \be
 F(x^{+})=\phi(x^{+})p(x^{+}),\qquad p(x^{+})^{2}=\partial_{+}f(x^{+})+\phi(x^{+})\partial_{+}\phi(x^{+}).
 \ee
The transformation in the right-moving sector is 
\bea
{x^{-}}^{\prime}&=&x^{-}+g(x^{+})+\theta G(x^{+}), 
\eea
where $G(x^+)$ is an anticommuting holomorphic function. Considering the infinitesimal version of the above transformations, we find that the generators of the super-Virasoro-Kac-Moody algebra  
could be realized 
 by 
\bea
L_n&=&(x^+)^{n+1}\p_{+}+\frac{1}{2}(n+1)(x^+)^{n}\th\p_{\th}, \nn\\
P_n&=&(x^+)^{n}\p_{-}, \nn\\
G_r&=&(x^+)^{r+1/2}(\p_\th-\th\p_{+}),\nn\\
S_s&=&-2(x^+)^{s-1/2}\th\p_{-}. 
\eea
They satisfy the above commutation relations without central extensions. Then from the Jacobi identity, the central extensions could be recovered. This fact shows that the chiral superspace $(x^+,\th)$ is enough for our study.

\section{Properties of SWCFT$_2$}

Now we have found two kinds of minimal theories in $N=(1,0)$ superspace, starting from a 2D QFT with chiral scaling and translation symmetry. One is the $N=(1,0)$ supersymmetric conformal field theory, whose local symmetries consist of a left-moving super-Virasoro algebra (SVA) and a right-moving Virasoro algebra. The other 
 is the supersymmetric warped conformal field theory, whose local symmetries are generated by supersymmetric Virasoro-Kac-Moody algebra (SVCMA). In this section, we discuss the representations of this algebra,  the state-operator correspondence and then the correlation functions in SWCFT$_2$. 
 %Super-Warped-Conformal Algebra(SWCA), which we mainly discussed in the following.

\subsection{Primary states and descendants}

In all our subsequent discussions, we consider the NS sector of SWCFT$_2$ and hence $r,s\in \mathbb{Z}+\frac{1}{2}$. We want to define the states in this theory at $t=0$ by doing radial quantization.   For this purpose, we consider the following complex coordinates
\bea
x^{+}=e^{-i(t-\phi)}=e^{i\phi+t_{E}},\qquad x^{-}=t+2\gamma (\phi-t).
\eea
where $t$ is interpreted as the Lorentzian time, and $t_{E}=-it$ as the Euclidean time. Having an initial state at very early Euclidean time corresponds to insert an operator at $x^{+} = 0$. Using translational symmetry, we can further put the operator at $x^{-}=0$. A primary operator $\Phi$ of  weight $\Delta$ and charge $Q$ at $x^{+} = 0$ corresponds to a state
\bea
\Phi(0,0,0)\sim|\Delta,Q>.
\eea
In particular, because of the global sub-algebra of SVKMA is $osp(1|2)\times u(1)$, the identity operator corresponds to the $OSP(1|2)\times U(1)$ invariant vacuum. The vacuum state $|0\rangle$ is defined as 
\bea
&\hs{1.0ex}L_{n}|0\rangle=0,\qquad\text{$n\ge -1$},\hs{0.0ex}\notag\\
&P_{m}|0\rangle=0,\qquad\text{$m\ge 0$},\hs{0.5ex}\notag\\
&\hs{1.2ex}G_{r}|0\rangle=0,\qquad\text{$r\ge -\frac{1}{2}$},\hs{0.0ex}\notag\\
&S_{s}|0\rangle=0,\qquad\text{$s\ge \frac{1}{2}$}.\hs{0.5ex}
\eea

We now construct the representations by considering the states having definite scaling dimensions and $U(1)$ charges. The state $|\Delta,Q\rangle$  is of scaling dimension $\D$ and charge $Q$ \bea
&L_{0}|\Delta,Q\rangle=\Delta|\Delta,Q\rangle,\notag\\
&P_{0}|\Delta,Q\rangle=Q|\Delta,Q\rangle.
\eea
Using the algebra obtained previously, we have
\bea
L_{0}L_{n}|\Delta,Q\rangle&=(\Delta-n)L_{n}|\Delta,Q\rangle,\notag\\
L_{0}P_{m}|\Delta,Q\rangle&=(\Delta-m)P_{m}|\Delta,Q\rangle,\notag\\
L_{0}G_{r}|\Delta,Q\rangle&=(\Delta-r)G_{r}|\Delta,Q\rangle,\notag\\
L_{0}S_{s}|\Delta,Q\rangle&=(\Delta-s)S_{s}|\Delta,Q\rangle.
\eea
We can see that the positive modes $L_{n}$, $P_{m}$, $G_{r}$, $S_{s}$ lower the value of the scaling dimension while the negative modes $L_{-n}$, $P_{-m}$, $G_{-r}$, $S_{-s}$ raise the value of the scaling dimension. The super-primary states in the theory are defined to have the following properties
\bea
 &&L_{n}|\Delta,Q\rangle=0,\quad  \notag\\
&&P_{n}|\Delta,Q\rangle=0,\quad \text{$n>0$},\notag\\
&&G_{r}|\Delta,Q\rangle=0,\quad\text{$r>0$},\notag\\
&&S_{s}|\Delta,Q\rangle=0,\quad\text{$s>0$}.
\eea

The  modules (analogue to the Verma modules in CFT$_2$) in SWCFT$_2$ are then defined by acting the raising operators $L_{-n}$, $P_{-m}$, $G_{-r}$, $S_{-s}$, $n,m,r,s>0$ on the primary states. The descendant states at level $N$ is 
\bea
|\Delta,Q,\{N\}\rangle=L_{-n_{1}}\cdots L_{-n_{k}}P_{-m_{1}}\cdots P_{-m_{l}}G_{-r_{1}}\cdots G_{-r_{i}}S_{-s_{1}}\cdots S_{-s_{j}}|\Delta,Q\rangle,
\eea
where $\{N\}$ denotes four sets of  $\{n\}$, $\{m\}$, $\{r\}$ and $\{s\}$ and the total level $N$ is the sum of all elements in the sets. A primary module consists of a primary state and all its descendant states. %All super-descendents together with the highest weight state, by construction carry a representation.

In our chiral superspace, the superfield has two component fields which are related to each other by supersymmetric transformation. The states corresponding to the component fields can be obtained from the highest weight state
\bea
&|\varphi\rangle=|\Delta,Q\rangle,\notag\\
&|\psi\rangle=G_{-\frac{1}{2}}|\Delta,Q\rangle.
\eea
They share the same $P_0$ charge
\bea
P_{0}|\varphi\rangle=Q|\varphi\rangle,\notag\\
P_{0}|\psi\rangle=Q|\psi\rangle.
\eea

The matrix of inner products of the states including the descendants defines the SWCFT analogue of the Kac matrix in CFT. We will denote it by $\mathscr{M}_{N}$ and its matrix elements are
\bea
\mathscr{M}_{\{N\},\{N^{\prime}\}}=\langle\Delta,Q,\{N\}|\Delta,Q,\{N^{\prime}\}\rangle.
\eea
At level $\frac{1}{2}$, we have
\bea
\mathscr{M}_{\frac{1}{2}}&=\begin{bmatrix}  \langle\Delta,Q,|G_{\frac{1}{2}}\\ \langle\Delta,Q,|S_{\frac{1}{2}}\end{bmatrix}[G_{-\frac{1}{2}}|\Delta,Q\rangle,S_{-\frac{1}{2}}|\Delta,Q\rangle] =\begin{bmatrix}  2\Delta &2Q\\2Q&\frac{k}{2}\end{bmatrix}.
\eea
At level 1, we have
\bea
\mathscr{M}_{1}&=&\begin{bmatrix} \langle\Delta,Q,|L_{1}\\ \langle\Delta,Q,|S_{\frac{1}{2}}G_{\frac{1}{2}}\\ \langle\Delta,Q,|P_{1}\end{bmatrix}[L_{-1}|\Delta,Q\rangle,G_{-\frac{1}{2}}S_{-\frac{1}{2}}|\Delta,Q\rangle,P_{-1}|\Delta,Q\rangle] \notag\\
&=&\begin{bmatrix}  2\Delta &2Q&Q\\2Q&\frac{(2\Delta+1)k-8Q^{2}}{2}&\frac{k}{4}\\Q&\frac{k}{4}&\frac{k}{2} \end{bmatrix}.
\eea
At level $\frac{3}{2}$, we have
\bea
\mathscr{M}_{\frac{3}{2}}&=\begin{bmatrix}  2\Delta(2\Delta+1)&4Q&(2\Delta+1)2Q&2\Delta Q &2Q&2Q^{2}\\4Q&2\Delta+\frac{2}{3}c&4Q&Q&2Q&\frac{k}{4}\\(2\Delta+1)2Q&4Q& \frac{(2\Delta+1)k}{2}&2Q^{2}&\frac{k}{2}&\frac{kQ}{2}\\2\Delta Q&Q&2Q^{2}&k\Delta &0&k\Delta\\2Q&2Q&\frac{k}{2}&0&\frac{k}{2}&0\\2Q^{2}&\frac{k}{4}&\frac{kQ}{2}&k\Delta&0&\frac{k^{2}}{4}\end{bmatrix},
\eea
which is in the base
 \be
 \{L_{-1}G_{-\frac{1}{2}}|\Delta,Q\rangle,G_{-\frac{3}{2}}|\Delta,Q\rangle,L_{-1}S_{-\frac{1}{2}}|\Delta,Q\rangle,P_{-1}G_{-\frac{1}{2}}|\Delta,Q\rangle,S_{-\frac{3}{2}}|\Delta,Q\rangle,P_{-1}S_{-\frac{1}{2}}|\Delta,Q\rangle\}. \nn\ee

We can derive some simple unitarity bounds on the plane charges by requiring the norm of the states to be non-negative. First, we have 
\bea
&||L_{n}|\Delta,Q||\ge 0\quad\Longrightarrow \Delta\ge 0,\quad c\ge 0,\\
&||P_{n}|\Delta,Q||\ge 0\quad\Longrightarrow  Q\in\mathbb{R},\quad k\ge 0.
\eea
Furthermore, the matrix $\mathscr{M}_{\frac{1}{2}}$ gives
\bea
k\Delta-4Q^{2}\ge 0\quad\Longrightarrow k\ge \frac{4Q^{2}}{\Delta}.
\eea

\subsection{Transformation laws of superprimary fields}

We now consider the transformation laws of the primary superfields. The local operator at position $(x^{+},x^{-},\theta)$ is related to the one at the origin by  the transformation
\begin{equation}
\Phi(\mathbf{z})\equiv \Phi(x^{+},x^{-},\theta)=U\Phi(0)U^{-1},\hs{3ex}\text{with $U=e^{x^{+}L_{-1}+\theta G_{-\frac{1}{2}}+x^{-}P_{0}}$. }
\end{equation}
Next we would like to find the explicit form of the commutator $[L_{n}, \Phi(\mathbf{z})] (n\geq 0)$ for a primary field $\Phi(\mathbf{z})$. First, we have
\begin{equation}
[L_{n},\Phi(\mathbf{z})]=U[U^{-1}L_{n}U,\Phi(0)]U^{-1}.
\end{equation}
Using the Baker-Campbell-Hausdorff (BCH) formula, we get
\bea
U^{-1}L_{n}U&=&\sum_{k=0}^{n+1}\frac{(n+1)!}{k!(n+1-k)!}[(x^{+})^{k}L_{n-k}+\frac{k}{2}(x^{+})^{k-1}\theta G_{n+\frac{1}{2}-k}],\notag\\
U^{-1}P_{m}U&=&\sum_{k=0}^{m}\frac{m!}{k!(m-k)!}[(x^{+})^{k}P_{m-k}+(x^{+})^{k-1}\theta S_{m+\frac{1}{2}-k}],\notag\\
U^{-1}G_{r}U&=&\sum_{k=0}^{r+\frac{1}{2}}\frac{(r+\frac{1}{2})!}{k!(r+\frac{1}{2}-k)!}(x^{+})^{k}G_{n-k}-\sum_{k=0}^{r+\frac{3}{2}}\frac{(r+\frac{1}{2})!}{k!(r+\frac{3}{2}-k)!}[2k(x^{+})^{k-1}\theta L_{r+\frac{1}{2}-k}],\notag\\
U^{-1}S_{s}U&=&\sum_{k=0}^{s-\frac{1}{2}}\frac{(s-\frac{1}{2})!}{k!(s-\frac{1}{2}-k)!}(x^{+})^{k}S_{s-k}-2\sum_{k=0}^{s+\frac{1}{2}}\frac{(s-\frac{1}{2})!}{k!(s+\frac{1}{2}-k)!}\theta (x^{+})^{k-1}P_{s+\frac{1}{2}-k}.
\eea
Then we obtain
\bea
&&[L_{n},\Phi(\mathbf{z})]=U[(x^{+})^{n+1}L_{-1}+\frac{n+1}{2}(x^{+})^{n}\theta G_{-\frac{1}{2}}+(n+1)(x^{+})^{n}L_{0},\Phi(0)]U^{-1},\quad \text{$n\geq -1$},\notag\\
&&[P_{m},\Phi(\mathbf{z})]=U[(x^{+})^{m}P_{0},\Phi(0)]U^{-1},\quad \text{$m\geq 0$},\notag\\
&&
[G_{r},\Phi(\mathbf{z})]=U[(x^{+})^{r+\frac{1}{2}}(G_{-\frac{1}{2}}-2\theta L_{-1})-2(r+\frac{1}{2})(x^{+})^{r-\frac{1}{2}}\theta L_{0},\Phi(0)]U^{-1},\quad\text{$r\geq -\frac{1}{2}$},\notag\\
&&
[S_{s},\Phi(\mathbf{z})]=(x^{+})^{s-\frac{1}{2}}U[-2\theta P_{0},\Phi(0)]U^{-1},\quad\text{$s>0$}.
\eea
In particular, we have
\bea
UL_{-1}U^{-1}=L_{-1},&
&UP_{0}U^{-1}=P_{0},\notag\\
UG_{-\frac{1}{2}}U^{-1}=G_{-\frac{1}{2}}+2\theta L_{-1},&
&US_{\frac{1}{2}}U^{-1}=S_{\frac{1}{2}}+2\theta P_{0}.
\eea
Using the above relations, we finally obtain 
\bea
\label{dif1}[L_{n},\Phi(\mathbf{z})]&=&[(x^{+})^{n+1}\partial_{+}+\frac{n+1}{2}(x^{+})^{n}\theta \partial_{\theta}
+(n+1)(x^{+})^{n}\Delta]\Phi(\mathbf{z}),\quad \text{$n\geq -1$},\\
\label{dif2}
[P_{m},\Phi(\mathbf{z})]&=&(x^{+})^{m}\partial_{-}\Phi(\mathbf{z}),\quad \text{$m\geq 0$},\\
\label{dif3}
[G_{r},\Phi(\mathbf{z})]&=&[(x^{+})^{r+\frac{1}{2}}(\partial_{\theta }-\theta \partial_{+})-2(r+\frac{1}{2})(x^{+})^{r-\frac{1}{2}}\theta \Delta]\Phi(\mathbf{z}),\quad\text{$r\geq -\frac{1}{2}$},\\
\label{dif4}
[S_{s},\Phi(\mathbf{z})]&=&-2(x^{+})^{s-\frac{1}{2}}\theta \partial_{-}\Phi(\mathbf{z}),\quad\text{$s>0$}.
\eea

%\noindent {\large{\bf Acknowledgments}} \\

\subsection{Ward identities and correlation functions}

The vacuum of the NS sector in SWCFT$_2$ is invariant under the global group $OSP(1|2)\times U(1)$, which is generated by $\{L_{0}, L_{\pm 1}, P_{0}, G_{\pm\frac{1}{2}}\}$. Thus the correlation functions  obey the Ward identities coming from the  generators of $OSP(1|2)\times U(1)$. One can solve the differential equations from the Ward identities to find the correlation functions directly using (\ref{dif1})-(\ref{dif4}). 

Consider an $n$-point function of super-primary fields 
\bea
G^{(n)}(\{\mathbf{z}_{i}\})&\equiv& G^{(n)}(\{x^{+}_{i},x^{-}_{i},\theta_{i}\})\notag\\
&=&\langle 0|T[ \Phi_1(x^{+}_{1},x^{-}_{1},\theta_{1})\Phi_2(x^{+}_{2},x^{-}_{2},\theta_{2})\cdots \Phi_n(x^{+}_{n},x^{-}_{n},\theta_{n})]|0\rangle,
\eea
where $T$ stands for time ordering, and $\mathbf{z}_{i}\equiv\{x^{+}_{i},x^{-}_{i},\theta_{i}\}$. Throughout this paper, we will always assume $x^{+}_{i}>x^{+}_{j}$ for $i<j$. Since the vacuum state $|0\rangle$ is $OSP(1|2)\times U(1)$ invariant, the $n$-point function is invariant
under the action of $L_{0}, L_{\pm 1}, G_{\pm\frac{1}{2}}$. This leads to the following differential equations corresponding to the generators $L_{-1}, P_0, L_0, G_{-1/2}, G_{1/2}$ respectively 
\bea
0&=&\left(\sum_{i=1}^{n}\partial_{i+}\right)G^{(n)},\notag\\
0&=&(\partial_{i-}-Q_{i})G^{(n)},\qquad \text{with} \hs{2ex} \sum_{i=1}^{n}Q_{i}=0,\notag\\
0&=&\left(\sum_{i=1}^{n}(x_{i}^{+}\partial_{i+}+\frac{1}{2}\theta_{i}\partial_{\theta_{i}}+\Delta_{i})\right)G^{(n)},\notag\\
0&=&\left(\sum_{i=1}^{n}((x_{i}^{+})^{2}\partial_{i+}+x_{i}^{+}\theta_{i}\partial_{\theta_{i}}2x_{i}^{+}\Delta_{i})\right)G^{(n)},\notag\\
0&=&\left(\sum_{i=1}^{n}\partial_{\theta_{i}}-\theta_{i}\partial_{i+}\right)G^{(n)},\notag\\
0&=&\left(\sum_{i=1}^{n}x_{i}^{+}(\partial_{\theta_{i}}-\theta_{i}\partial_{i+})-2\theta_{i}\Delta)\right)G^{(n)}.
\eea
 The first equation implies that $G^{(n)}$ should be a function of $x_{ij}^{+}\equiv x_{i}^{+}-x_{j}^{+}$. While the fourth equation implies that $G^{(n)}$ should be a function of
\begin{equation}
s_{ij}\equiv x_{i}^{+}-x_{j}^{+}-\theta_{i}\theta_{j},
\end{equation}
with
\begin{equation}
\theta_{ij}\equiv \theta_{i}-\theta_{j},
\end{equation}
For the $x^{-}$ part, the second equation implies $G^{(n)}$ should be a function of
\bea
r_{ij}\equiv x_{i}^{-}Q_{i}+x_{j}^{-}Q_{j}.
\eea
Consequently, the  correlation function is of form
\begin{equation}
G^{(n)}(s_{ij},r_{ij},\theta^{+}_{ij}).
\end{equation}
We stress that because of the $OSP(1|2)\times U(1)$ symmetry of the vacuum, the correlation functions must have the $OSP(1|2)\times U(1)$ structure.

%\subsubsection{Two-point functions}

Let us first consider the two-point function $G^{(2)}$ which must be of form
\be
G^{(2)}(\mathbf{z_{1}},\mathbf{z_{2}})=\frac{f_{1}(r_{12})}{(s_{12})^{\kappa_{1}}}
+\frac{f_{2}(r_{12})\theta_{12}}{(s_{12})^{\kappa_{2}}},
\ee
where $\{f_{i}\}$ are the functions to be determined. The equation from the invariance under $P_{0}$ gives
\bea
\frac{\partial f_{i}}{\partial r_{12}}=f_{i}, \hs{3ex}
Q_{1}+Q_{2}=0, 
\eea
which has the solution
\bea
f_{i}=C_{i}e^{r_{12}},\hs{3ex} i=1,2
\eea
where $\{C_{i}\}$ are constants. 

Next consider  the differential equation arising from dilations $L_{0}$, we  find the conditions
\bea
&-\kappa_{1}+\Delta_{1}+\Delta_{2}=0,\notag\\
&-\kappa_{2}+\frac{1}{2}+\Delta_{1}+\Delta_{2}=0.
\eea
Moreover, the special transformation $L_{+1}$ leads to the condition
\bea
0&=C_{1}(s_{12})^{\frac{1}{2}}\left[\Delta_{12}x^{+}_{12}\right]+C_{2}\left[(\Delta_{12}+\frac{1}{2}) x^{+}_{12}\theta_{1}-(\Delta_{12}-\frac{1}{2}) x^{+}_{12}\theta_{2}\right],\notag\\
\eea
which gives
\bea
\Delta_{12}=0,\qquad C_{2}=0.
\eea
As a result, we have the two-point function
\bea
G^{(2)}(\mathbf{z_{1}},\mathbf{z_{2}})&=\delta_{\Delta_{1},\Delta_{2}}\delta_{Q_{1},-Q_{2}}\frac{1}{s_{12}^{2\Delta_{1}}}e^{x^{-}_{12}Q_{1}},
\eea
where we have set the normalization to unit. In the component fields, the nonvanishing two-point functions are 
\bea
&\langle \varphi_1 \varphi_2\rangle=\delta_{\Delta_{1},\Delta_{2}}\delta_{Q_{1},-Q_{2}}\frac{1}{(x^{+}_{12})^{2\Delta_{1}}}e^{x^{-}_{12}Q_{1}},\notag\\
&\langle \psi_1\psi_2 \rangle=\delta_{\Delta_{1},\Delta_{2}}\delta_{Q_{1},-Q_{2}}\frac{2\Delta_{1} }{(x^{+}_{12})^{2\Delta_{1}+1}}e^{x^{-}_{12}Q_{1}}, 
\eea
where $\varphi_i $ and $\psi_i$ are the component fields of the superfield $\Phi_i$. 

It is clear that the two-point functions respect the $OSP(1|2)\times U(1)$ symmetry. The $OSP(1|2)$ part is determined by the modes $\{L_{0}, L_{\pm 1}, G_{\pm\frac{1}{2}}\}$, just as the usual $N=(1,0)$ SCFT$_2$, while the $U(1)$ part is totally determined by the zero mode $P_{0}$. Moreover, we note that the two-point-functions of the  superprimaries in SWCFT$_2$ could be  obtained straightforwardly from the bosonic one \cite{Song:2017czq} by replacing the difference of the two bosonic coordinates $x^{+}_{12}$ with its supersymmetric generalization $s_{12}$.

%\subsubsection{Higher-point functions}

From the structures of two-point function, we know that the higher-point correlation functions of SWCFT$_2$ must also include two parts, one being determined by the modes $\{L_{0}, L_{\pm 1}, G_{\pm\frac{1}{2}}\}$, the other being determined by the zero mode $P_{0}$. We can use the well known results of $N=1$ super conformal theory to determine the $OSP(1|2)$ structures of the correlation functions of SWCFT$_2$. For example, the holomorphic three-point function in the NS sector of $N=1$ super-conformal theory is given in\cite{Qiu:1986if}, and the $U(1)$ part is given by \cite{Song:2017czq}, thus the three-point function in the NS sector of the SWCFT primaries is
\bea
G^{(3)}(\mathbf{z_{1}},\mathbf{z_{2}},\mathbf{z_{3}})=\delta_{Q_{1}+Q_{2}+Q_{3},0}\frac{C_{123}+\tilde{C}_{123}\varXi_{123}}{s_{12}^{\Delta_{1}+\Delta_{2}-\Delta_{3}}s_{31}^{\Delta_{3}+\Delta_{1}-\Delta_{2}}s_{23}^{\Delta_{2}+\Delta_{3}-\Delta_{1}}}e^{\frac{1}{3}Q_{12}x^{-}_{12}}e^{\frac{1}{3}Q_{31}x^{-}_{31}}e^{\frac{1}{3}Q_{23}x^{-}_{23}}, \nn\\
\eea
where we have defined $x^{-}_{ij}\equiv x^{-}_{i}-x^{-}_{j}$, $Q_{ij}\equiv Q_{i}-Q_{j}$ and $\Delta_{ijk}\equiv\Delta_{i}+\Delta_{j}-\Delta_{k}$. The $C_{123}$ and $\tilde{C}_{123}$ are two structure constants of the three-point function. The quantity $\varXi_{ijk}$ is given by
\bea
\varXi_{ijk}\equiv\frac{s_{ij}\theta_{k}+s_{jk}\theta_{i}+s_{ki}\theta_{j}}{\sqrt{s_{ij}s_{jk}s_{ki}}}.
\eea

For the $n$-point function ($n>3$) there are $3n$ coordinates $\{x^{+}_{i},x^{-}_{i},\theta_{i}\}$, and 5
constraints from $OSP(1|2)$ invariance and one constraint from $U(1)$ invariance.  The $x^{-}_{i}$ dependence can be totally determined by the $U(1)$ invariance, thus the $n$-point function is  essentially a function of $(2n-5)$ $OSP(1|2)$ invariants, which are given by\cite{Fuchs:1986ew}
\be
\varXi_{ijk}, \hs{3ex} \Theta_{ijkl}\equiv\frac{s_{ij}s_{kl}}{s_{li}s_{jk}}.
\ee
The general form of $n$-point function can be written as
\bea
G^{(n)}(\{\mathbf{z}_{i}\})=\delta_{\sum_{i=1}^{n}Q_{i},0}\left(\prod_{i<j=1}^{n}e^{\frac{r_{ij}}{4}}\right)\left(\prod_{i<j=1}^{n}s_{ij}^{-\Delta_{ij}}\right)F(\varXi_{ijk},\Theta_{ijkl}). 
\eea
Here $F(\varXi_{ijk},\Theta_{ijkl})$ is an undetermined function, and $\Delta_{ij}$ are real constants which satisfy
\bea
\sum_{i\ne j}\Delta_{ij}=2\Delta_{j},\qquad \Delta_{ij}=\Delta_{ji}.
\eea

%\noindent {\large{\bf Acknowledgments}} \\

\section{Conclusion and Discussion}

In the present work we studied supersymmetric extension of the warped conformal field theory. Under the assumption that the dilation operator is diagonalizable, and has a discrete, non-negative spectrum, we generalized the Hofman-Strominger theorem to the supersymmetric case. Specifically, we showed that  a two-dimensional quantum field theory with two translational symmetries, a chiral  scaling symmetry and a chiral supersymmetry may have enhanced local symmetry. The global symmetry could be enhanced to  two kinds of  minimal algebra. One consists of one copy of the Virasoro algebra and one copy of the super-Virasoro algebra, which leads to the $N=(1,0)$ SCFT$_{2}$. The other consists of  one copy of the super-Virasoro algebra plus a $U(1)$ super-Kac-Moody algebra, which leads to the  $N=1$ supersymmetric warped conformal field theory. 

We discussed some properties of the SWCFT$_2$, including the representations of the algebra, the space of the states and the transformations of the superfields. We furthermore  calculated the two-point and three-point correlation functions of the  SWCFT$_2$ with the help of chiral superspace.  The form of the correlation functions can be fixed without involving a specific model. Particularly, the vacuum of NS sector in SWCFT$_2$ is $OSP(1|2)\times U(1)$ invariant such that the correlation functions  must have the $OSP(1|2)\times U(1)$ structure, in which the $U(1)$ symmetry determined the dependence on $x^-$ completely.

One possible future direction is to generalize the minimal supersymmetry to the extended one.  
Our construction is based on the chiral superspace $(x^+,\th)$. It is worthy of generalizing the study to the full superspace, including the Grassmannian partner of the $x^-$ coordinate.  In the minimal CFT$_2$ case, this may lead to the $N=(1,1)$ SCFT$_2$. But it is not clear of its consequence in the WCFT$_2$ case.  The study can be pushed to the case of $N\geq 2$ extended supersymmetry as well. Besides, it is interesting to study the supersymmetrization of the other 2D models with scaling symmetry. The supersymmetric GCA has been studied in \cite{Mandal:2010gx}, but for more general AGFT\cite{Chen:2019hbj} its supersymmetric version has not been worked out. 

It would be interesting to study the other properties of SWCFT$_2$: the modular properties of the torus partition function, the warped conformal bootstrap\cite{Song:2017czq,Apolo:2018eky}, the entanglement entropy, etc.. It is also interesting to construct explicitly simple examples realizing the SWCFT$_2$. This may help us to understand the theory better.

It could be expected that for the holographic SWCFT$_2$, it is dual to a supersymmetric  AdS$_3$ gravity under appropriate asymptotic boundary conditions. It would be nice to find the explicit boundary conditions and see how they break half of the supersymmetries.

\vspace*{.5cm}
\noindent {\large{\bf Acknowledgments}} \\

The work  was in part supported by NSFC Grant No.~11275010, No.~11335012, No.~11325522 and No. 11735001. 

\newpage
\section*{Appendix: conserved charges in the superspace}
We start from the global symmetries of the theory. It is generated by  the left-moving  translation $H$, the dilation $D$, the right-moving translation $\overline{P}$ and the supersymmetries $Q_{+}$ and $Q_{-}$. By assumption these charges annihilate the vacuum. Their non-vanishing  (anti)commutation relations are
\begin{equation}\label{charge}
i\big\{Q_{+},Q_{+}\big\}=2H,\qquad i\big\{Q_{+},Q_{-}\big\}=2\overline{P},\qquad i\big\{Q_{-},Q_{-}\big\}=0,
\end{equation}
\begin{equation}\label{diemnsion1}
i[D,H]=H, \qquad\qquad i[D,\overline{P}]=0,           
\end{equation}
\begin{equation}\label{diemnsion2}
i[D,Q_{+}]=\frac{1}{2}Q_{+},\qquad i[D,Q_{-}]=-\frac{1}{2}Q_{-}.                                  
\end{equation}

The superspace is a coset space $G/I$, where $G$ is the whole symmetry group and  $I$ is the dilation symmetry. A group element in $G$ may be written in the form
\begin{equation}
g_{0}=e^{i(\delta H+\epsilon^{+}Q_{+}+\overline{\delta}\overline{P}+\epsilon^{-}Q_{-})}e^{i\lambda D},
\end{equation}
where $\delta$, $\overline{\delta}$, $\epsilon^{+}$, $\epsilon^{-}$ and $\lambda$ are some infinitesimal constants. The coset element can be written as
\begin{equation}
g_{1}=e^{i(x^{+}H+\theta^{+}Q_{+}+x^{-}\overline{P}+\theta^{-}Q_{-})}.
\end{equation}
The transformations on the superspace are the natural action of the group $G$ on the coset space
\begin{equation}
g_{0}g_{1}=e^{i({x^{+}}^{\prime}H+{\theta^{+}}^{\prime}Q_{+}+{x^{-}}^{\prime}\overline{P}+{\theta^{-}}^{\prime}Q_{-})}e^{i\lambda^{\prime}D},
\end{equation}
from which we read the induced transformations in the superspace
\bea
{x^{+}}^{\prime}&=&x^{+}+\delta+\lambda x^{+}-\epsilon^{+}\theta^{+},\\
{\theta^{+}}^{\prime}&=&\theta^{+}+\epsilon^{+}+\frac{\lambda}{2}\theta^{+},\\
{\theta^{-}}^{\prime}&=&\theta^{-}+\epsilon^{-}-\frac{\lambda}{2}\theta^{-},\\
{x^{-}}^{\prime}&=&x^{-}+\overline{\delta}-\epsilon^{-}\theta^{+}-\epsilon^{+}\theta^{-}.
\eea
Then we can obtain the differential representations of the global charges
\bea
H&=&-i\partial_{+},\hs{4ex}\overline{P}=-i\partial_{-},\notag\\
Q_{+}&=&-i(\partial_{\theta^{+}}-\theta^{+}\partial_{+}-\theta^{-}\partial_{-}),\notag\\
Q_{-}&=&-i(\partial_{\theta^{-}}-\theta^{+}\partial_{-}),\notag\\
D&=&-i(x^{+}\partial_{+}+\frac{\theta^{+}}{2}\partial_{\theta^{+}}-\frac{\theta^{-}}{2}\partial_{\theta^{-}}).
\eea

For each of the charges $H$, $D$, $\overline{P}$, $Q^{+}$ and $Q^{-}$, there is a conserved Noether current. In particular, with the supersymmetries, there exist corresponding supercurrents. In general the supercurrents may have the form 
\bea
\mathbb{O}_{+}(x^{+},x^{-},\theta^{+},\theta^{-})&=a_{1}{o}_{+1}+a_{2}\theta^{+} {o}_{+2}+a_{3}\theta^{-} {o}_{+3}+a_{4}\theta^{+}\theta^{-} {o}_{+4},\notag\\
\mathbb{O}_{-}(x^{+},x^{-},\theta^{+},\theta^{-})&=a_{1}{o}_{-1}+a_{2}\theta^{+} {o}_{-2}+a_{3}\theta^{-} {o}_{-3}+a_{4}\theta^{+}\theta^{-} {o}_{-4},
\eea
where $a_{i}$ ($i=1, 2, 3, 4$) are constant numbers. The charges associated to the components of supercurrents can be read by
\bea
O_{i}&=-\frac{1}{2\pi}\int dx^{+}{o}_{+i}+\frac{1}{2\pi}\int dx^{-}{o}_{-i}.
\eea
The supersymmetric transformations of the supercurrents  are 
\bea
\lefteqn{i[\epsilon_{1} Q_{+},\mathbb{O}_{+}(x^{+},x^{-},\theta^{+},\theta^{-})]}\notag\\
&=&\epsilon_{1}(\partial_{\theta^{+}}-\theta^{+}\partial_{+}-\theta^{-}\partial_{-})(a_{1}{o}_{+1}+a_{2}\theta^{+} {o}_{+2}+a_{3}\theta^{-} {o}_{+3}+a_{4}\theta^{+}\theta^{-} {o}_{+4}),\notag\\
&=&\epsilon_{1}\left( a_{2}{o}_{+2}-\theta^{+}a_{1}\partial_{+}{o}_{+1}+\theta^{-}(a_{4}{o}_{+4}-a_{1}\partial_{+}{o}_{+1})+\theta^{+}\theta^{-}(a_{2}\partial_{-}{o}_{+2}-a_{3}\partial_{+}{o}_{+3})\right),\nn\\
& & \\
\lefteqn{i[\epsilon_{2} Q_{-},\mathbb{O}_{+}(x^{+},x^{-},\theta^{+},\theta^{-})]}\notag\\
&=&\epsilon_{2}(\partial_{\theta^{-}}-\theta^{+}\partial_{-})(a_{1}{o}_{+1}+a_{2}\theta^{+} {o}_{+2}+a_{3}\theta^{-} {o}_{+3}+a_{4}\theta^{+}\theta^{-} {o}_{+4}),\notag\\
&=&\epsilon_{2}\left(a_{3} {o}_{+3}-\theta^{+}(a_{4}{o}_{+4}+a_{1}\partial_{-}{o}_{+1})-a_{3}\theta^{+}\theta^{-}\partial_{+}{o}_{+3}\right),
\eea
where $\epsilon_{i}$($i=1,2$) are the Grassmannian constants. Then we have 
\bea
i[\epsilon_{1} Q_{+},a_{1}{o}_{+1}]&=&\epsilon_{1}a_{2}{o}_{+2},\notag\\
i[\epsilon_{1} Q_{+},a_{2}{o}_{+2}]&=&-\epsilon_{1}a_{1}\partial_{+}{o}_{+1},\notag\\
i[\epsilon_{1} Q_{+},a_{3}{o}_{+3}]&=&\epsilon_{1}(a_{4}{o}_{+4}-a_{1}\partial_{+}{o}_{+1}),\notag\\
i[\epsilon_{1} Q_{+},a_{4}{o}_{+4}]&=&\epsilon_{1}(a_{2}\partial_{-}{o}_{+2}-a_{3}\partial_{+}{o}_{+3}),
\eea
and 
\bea
i[\epsilon_{2} Q_{-},a_{1}{o}_{+1}]&=&\epsilon_{2}a_{3} {o}_{+3},\notag\\
i[\epsilon_{2} Q_{-},a_{2}{o}_{+2}]&=&-\epsilon_{2}(a_{4}{o}_{+4}+a_{1}\partial_{-}{o}_{+1}),\notag\\
i[\epsilon_{2} Q_{-},a_{3}{o}_{+3}]&=&0,\notag\\
i[\epsilon_{2} Q_{-},a_{4}{o}_{+4}]&=&-\epsilon_{2}a_{3}\partial_{+}{o}_{+3},
\eea
and similarly for $\mathbb{O}_{-}(x^{+},x^{-},\theta^{+},\theta^{-})$. After the integration, we get the transformations of the charges associated to the components of supercurrents
\bea
i[\epsilon_{1} Q_{+},a_{1}{O}_{1}]=\epsilon_{1}a_{2}{O}_{2}, 
&i[\epsilon_{1} Q_{+},a_{2}{O}_{2}]=0,\notag\\
i[\epsilon_{1} Q_{+},a_{3}{O}_{3}]=\epsilon_{1}a_{4}{O}_{4},
&i[\epsilon_{1} Q_{+},a_{4}{O}_{4}]=0,\label{trans1}\\
i[\epsilon_{2} Q_{-},a_{1}{O}_{1}]=\epsilon_{2}a_{3} {O}_{3},
&\hs{7.5ex}i[\epsilon_{2} Q_{-},a_{2}{O}_{2}]=-\epsilon_{2}a_{4}{O}_{4},\notag\\
i[\epsilon_{2} Q_{-},a_{3}{O}_{3}]=0,
\hs{5.8 ex}&i[\epsilon_{2} Q_{-},a_{4}{O}_{4}]=0.\label{trans2}
\eea

In the following, we will donate the currents associated to the charges $H$, $\overline{P}$, $Q_{+}$ and $Q_{-}$ by $h_{\pm}$, $p_{\pm}$, ${q_{+}}_{\pm}$ and ${q_{-}}_{\pm}$, respectively. As it is not clear at this moment how these currents are related to each other by the supersymmetries, we first assume each of them belongs to a supercurrent donated by $\mathbb{H}_{\pm}$, $\mathbb{\overline{P}}_{\pm}$, $\mathbb{Q_{+}}_{\pm}$ and $\mathbb{Q_{-}}_{\pm}$, then we will find out the relationship between the currents by their transformations under the supersymmetries. For $\mathbb{Q_{+}}_{\pm}$, in order to be consistent with (\ref{trans1}) and (\ref{trans2}) it must satisfy
\bea
i[\epsilon_{1} Q_{+},a_{1}{Q_{+}}_{1}]=\epsilon_{1}a_{2}{Q_{+}}_{2},
&i[\epsilon_{1} Q_{+},a_{2}{Q_{+}}_{2}]=0,\notag\\
i[\epsilon_{1} Q_{+},a_{3}{Q_{+}}_{3}]=\epsilon_{1}a_{4}{Q_{+}}_{4},
&i[\epsilon_{1} Q_{+},a_{4}{Q_{+}}_{4}]=0,
\eea
and
\bea
i[\epsilon_{2} Q_{-},a_{1}{Q_{+}}_{1}]=\epsilon_{2}a_{3} {Q_{+}}_{3},
&\hs{3ex}i[\epsilon_{2} Q_{-},a_{2}{Q_{+}}_{2}]=-a_{4}\epsilon_{2}{Q_{+}}_{4},\notag\\
i[\epsilon_{2} Q_{-},a_{3}{Q_{+}}_{3}]=0,\hs{7.5ex}
&i[\epsilon_{2} Q_{-},a_{4}{Q_{+}}_{4}]=0. \hs{6.5ex}
\eea
We have  similar relations for $\mathbb{H}_{\pm}$, $\mathbb{\overline{P}}_{\pm}$, and $\mathbb{Q_{-}}_{\pm}$. From these relations, we find that the form of $\mathbb{{Q_{+}}}$ can only be
\bea
\mathbb{{Q_{+}}}_{\pm}=a_{1}{q_{+}}_{\pm}+a_{2}\theta^{+}h_{\pm}+a_{3}\theta^{-}p_{\pm},
\eea
and the form of $\mathbb{{Q_{-}}}$ must be
\bea
\mathbb{{Q_{-}}}_{\pm}=a_{1} {q_{-}}_{\pm}+a_{2}\theta^{+}p_{\pm},
\eea 
with all none-zero coefficients $a_{i}$,  for $i=1,2,3$.  We see that the $\overline{P}$ belongs to two different supermultiplets. On the other hand, the fact that the operator $Q_-$ is nilpotent indicates that we may consider a smaller superspace. In fact we  can regard $Q_{-}$ as  the superpartner of $\overline{P}$, and consider only one global supercharge $Q_{+}$ in the theory. 
It turns out that the smaller superspace $\{x^{+}, x^{-}, \theta^{+}\}$ is enough to describe our theories consistently. 

The superspace $\{x^{+}, x^{-}, \theta^{+}\}$ is the coset space $G/\tilde{I}$, where  $G$ is the whole symmetry group and  $\tilde{I}$ consists of the dilation symmetry and $Q_{-}$. In this smaller superspace, we have
\bea
a_{3}=a_{4}=0,\qquad a_{2}=2a_{1},
\eea
and
\bea
\mathbb{H}_{\pm}=\mathbb{{Q_{+}}}_{\pm},\qquad \mathbb{\overline{P}}_{\pm}=\mathbb{{Q_{-}}}_{\pm}.
\eea
We may choose $a_{1}=1$, and find that 
\bea
&\mathbb{H}_{\pm}=\mathbb{{Q_{+}}}_{\pm}=h_{0\pm}(x^{+},x^{-})+2\theta^{+} h_{1\pm}(x^{+},x^{-}),\\
&\mathbb{\overline{P}}_{\pm}=\mathbb{{Q_{-}}}_{\pm}=p_{0\pm}(x^{+},x^{-})+2\theta^{+} p_{1\pm}(x^{+},x^{-}).
\eea
%For simplicity, we will always donate $\theta^{+}$ by $\theta$ in the  discussions.

\end{document}